\documentclass[11pt,titlepage]{article}
\usepackage[dvips]{graphics,epsfig}
\usepackage[margin=70pt]{geometry}

\usepackage{amsmath}
\usepackage{amssymb,amsfonts}

 \newtheorem{theorem}{Theorem}
 \newtheorem{lemma}{Lemma}
 \newtheorem{corollary}{Corollary}

 \newtheorem{claim}{Claim}

\newtheorem{remark}{Remark}
\newenvironment{proof}{%
  \noindent{\it Proof\ }}{%
  \hspace*{\fill}$\blacksquare$
  \vspace{2ex}\\}

 \newcommand{\ket}[1]{| #1 \rangle} \newcommand{\bracket}[2]{\langle #1 | #2 \rangle}

\newcommand{\calF}{{\cal F}}

\newcommand{\email}[1]{\texttt{#1}}

\newcommand{\set}[1]{\{ #1 \}}
\newcommand{\floor}[1]{\lfloor #1 \rfloor}
\newcommand{\Ham}{\mathrm{Ham}}

\begin{document}
\title{Average/Worst-Case Gap of Quantum Query Complexities\\ by On-Set Size}

\author{
Andris Ambainis\thanks{\scriptsize
Institute of Mathematics and Computer Science, University of Latvia, Latvia.
\email{ambainis@lu.lv}.\
} 
\and 
Kazuo Iwama\thanks{\scriptsize
School of Informatics, Kyoto University.
Kyoto, 
Japan.
\email{iwama@kuis.kyoto-u.ac.jp}.
} 
\and  
Masaki Nakanishi\thanks{\scriptsize
Faculty of Education, Art and Science, Yamagata University. Yamagata,
Japan.
\email{m-naka@e.yamagata-u.ac.jp}.
}
 \and 
Harumichi Nishimura\thanks{\scriptsize
School of Science, Osaka Prefecture University.
Osaka, 
Japan.
\email{hnishimura@mi.s.osakafu-u.ac.jp}.
}
\and 
Rudy Raymond\thanks{\scriptsize
Tokyo Research Laboratory, IBM Japan.
Kanagawa, 
Japan.
\email{raymond@jp.ibm.com}.
} 
\and 
Seiichiro Tani\thanks{\scriptsize
NTT Communication Science Laboratories, NTT Corporation.
Atsugi, 
Japan.
\email{tani@theory.brl.ntt.co.jp}.
}\thanks{\scriptsize
Quantum Computation and Information Project,
SORST, JST,
Tokyo, 
Japan.
}
\and
Shigeru Yamashita\thanks{\scriptsize
College of Information Science and Engineering, Ritsumeikan University.
\email{ger@cs.ritsumei.ac.jp}
}
}
\date{}
\maketitle

\begin{abstract}
This paper considers the query complexity of the functions in
the family $\calF _{N,M}$ of $N$-variable Boolean functions with
onset size $M$, i.e.,
the number of inputs for
which the function value is 1, where $1\le M \le 2^{N}/2$ is assumed without loss of generality
because of the symmetry of function values, 0 and 1.
Our main results are as follows: 
\begin{itemize}
\item There is a super-linear gap
between the average-case and worst-case quantum query complexities
over $\calF _{N,M}$ for a certain range of $M$.
\item There is no super-linear gap between 
the average-case and worst-case randomized query complexities
over $\calF _{N,M}$ for every $M$.
\item For every $M$ bounded by a polynomial in $N$, any function in $\calF_{N,M}$ has quantum query complexity $\Theta (\sqrt{N})$.
\item For every $M=O(2^{cN})$ with an arbitrary large constant $c<1$, any function in $\calF_{N,M}$ has randomized query complexity $\Omega (N)$.
\end{itemize}
\end{abstract}

\section{Introduction}
\subsection{Background}
Query complexities of Boolean functions are one of the most
fundamental and popular topics in quantum computation.  It is well
known that a quadratic speed-up, i.e., randomized query complexity $\Omega(N)$ to
quantum query complexity $O(\sqrt{N})$, is possible for several $N$-variable Boolean
functions including OR, AND, AND-OR trees (e.g.,
Refs.~\cite{Gro96STOC,HoyMosWol03ICALP,FarGolGut08TC,AmbChiReiSpaZha07FOCS}). However, we can obtain only a constant-factor speed-up (i.e.,
$\Omega(N)$ queries are needed in both classical and quantum settings) for other
Boolean functions such as PARITY~\cite{BeaBuhCleMosWol01JACM}.
Moreover, threshold functions
have
quantum query complexity depending on their
thresholds~\cite{BeaBuhCleMosWol01JACM}.
Thus we know well  about 
the quantum query complexity for Boolean functions
for these typical cases,
but much less is known
for the others.
Some known general results are
the worst-case and average-case query complexities 
(including the coefficients of dominant factors)
over all Boolean functions
in Refs.~\cite{Dam98FOCS} and~\cite{Amb99IPL}, respectively.
To understand more about the query complexity of all Boolean functions,
this paper examines
the query complexity for the set of Boolean functions with \emph{on-set} size $M$, 
i.e., with $M$ 1's on their truth tables, 
for every $M$.
Our results show that the  size of the
on-set of a Boolean function $f$ plays a key role in the query complexity of $f$, i.e.,
on-set size non-trivially bounds the quantum/randomized query complexity of $f$.
For instance, the quantum query complexity of \emph{every} function
with on-set size bounded by a polynomial in $N$ is $\Theta(\sqrt{N})$ 
while the randomized query complexity of the function is $\Omega(N)$,
as will be described later. 

The difference between average-case and worst-case complexities
is one of the central topics in theoretical computer science,
and it has been extensively studied for decades (e.g., Refs.~\cite{Pip77MST,AroBar09Book}).
However, in the quantum setting, only a few results are known.
(i) For a MAJORITY function, there is an almost quadratic gap between the
average-case and worst-case quantum query complexities over all inputs of the
function~~\cite{BeaBuhCleMosWol01JACM,AmbWol01JPA}. 
(ii) If we consider the 
average-case and 
worst-case behaviors of
complexities over all Boolean functions (for the
worst input of each function), only a linear gap is possible for quantum query
complexity~\cite{Dam98FOCS,Amb99IPL,ODonSer08JCSS}
and exact quantum communication complexity~\cite{buhrman-dewolf01}.
Our results imply 
a super-linear tight gap
between the average-case and worst-case \emph{quantum} query complexities
over the family of Boolean functions with on-set size $M$ for 
every $M$ in a certain range.
In contrast, the gap between
the average-case and worst-case \emph{randomized} query complexities
is at most linear for any on-set size $M$, which is also an implication of our results.

\paragraph{Previous Work}
The research on quantum query complexity
started with the Deutcsh-Jozsa algorithm~\cite{DeuJoz92RSLA}
and
other algorithms for computing partial functions (e.g., Simon's algorithm~\cite{Sim97SICOMP}), followed by
Grover's quantum
search algorithm~\cite{Gro96STOC}, which also computes the
Boolean OR function of $N$ variables with $O(\sqrt{N})$
queries.  Since then, 
numerous results
have extensively appeared
in the literature, showing that similar speed-ups are possible for many other
Boolean functions.
For example, if a Boolean function is given by a constant-depth 
balanced AND-OR trees (e.g., OR is by a single-depth tree), 
it can be computed in $O(\sqrt{N})$ quantum queries
with the robust quantum search technique~\cite{HoyMosWol03ICALP}.
This was recently extended to any AND-OR tree with
$O(N^{\frac{1}{2}+o(1)})$ quantum queries 
(optimal $O(\sqrt{N})$ quantum queries for nearly-balanced trees)
by using the quantum walk
technique~\cite{FarGolGut08TC,AmbChiReiSpaZha07FOCS}.
In general, however, 
the worst-case
quantum query complexity is polynomially related to 
the worst-case randomized query complexity for any Boolean function~\cite{BeaBuhCleMosWol01JACM}.
In contrast, there is an exponential gap between
the average-case randomized and quantum query complexities of a certain
Boolean function
for uniform distribution of inputs,
and the gap can be even larger for non-uniform distribution of inputs~\cite{AmbWol01JPA}.
As for the gap between the average-case and worst-case quantum query
complexities,
they are
$O(N^{1/2+\epsilon})$~\cite{AmbWol01JPA} and $\Omega
(N)$~\cite{BeaBuhCleMosWol01JACM}, respectively,  over all inputs for MAJORITY functions.
The average of complexity over all Boolean functions (for the worst input of each function)
was proved to be at least
$N/4-2\sqrt{N}\log N$~\cite{Amb99IPL}, which was improved to
$N/4+\Omega(\sqrt{N})$~\cite{ODonSer08JCSS},
and the worst-case complexity is
at most $N/2+\sqrt{N}$~\cite{Dam98FOCS}, respectively. 

In the circuit complexity theory,
it is known that
the maximum circuit size over the circuits 
for the family of Boolean functions with on-set size $M$
is closely related to binary entropy function
$H(p)$ for $p=M/2^N$ (e.g., Ref.~\cite{Pip77MST}).

\subsection{Our Results}  

\begin{table}[t]
  \centering
  \begin{tabular}{|c|| c|c|c|}\hline
    &Worst & Best & Average (Almost All)\\\hline\hline
Quantum &
$\Theta\left(\sqrt{N\frac{\log{M}}{c+\log{N}-\log\log{M}}}+\sqrt{N}\right)$
($\dagger$)
 &
$\Omega(\sqrt{N-\log M})$&
$ \Theta\left(\frac{\log{M}}{c + \log{N} -  \log\log{M}} + \sqrt{N}\right)$ \\\hline
Randomized&
$\Omega(N)$ &
$\Omega (N-\log M)$&
$\Omega (N)$\\\hline
  \end{tabular}
  \caption{Query Complexities of $N$-variable Boolean Functions with On-set Size $M$: ($\dagger$) holds for every $1 \le M \le 2^{N/(\log{N})^{2+\epsilon}}$ with an arbitrary
small positive constant $\epsilon$. The other bounds hold for every $1\le M \le 2^N/2$.}
  \label{tab:ResultSummary}
\end{table}

Let $\mathcal{F}_{N,M}$ be a family of $N$-variable Boolean functions $f_N$ 
with on-set size $M$, 
i.e.,
$f_N$ 
that have value $1$ (true) for $M$
assignments  in $\set{0,1}^N$.
Without loss of generality, we assume $M\in \{ 1,2,\dots ,2^N/2 \}$ because of the symmetry of function values, 0 and 1.
Let $Q(f_N)$ be the bounded-error quantum query complexity of $f_N$,
i.e., the number of quantum queries necessary to compute $f_N$
with bounded error
for the worst-case  input of $N$ bits given as an oracle.
We then investigate
the asymptotic behaviors of the following three functions of $N$ and $M$:
\begin{enumerate}
\item $Q_{\operatorname{worst}}(\calF _{N,M})\equiv \max _{f_N\in \calF _{N,M}} Q(f_N).$
\item $Q_{\operatorname{best}}(\calF _{N,M})\equiv \min _{f_N\in \calF _{N,M}} Q(f_N).$
\item $Q_{\operatorname{almost}}(\calF_{N,M})$ is an arbitrary function such that,
for uniformly distributed $f_N$ over $\mathcal{F}_{N,M}$, 
$\Pr_{f_N \in \mathcal{F}_{N,M}}[Q(f_N)=\Theta(Q_{\operatorname{almost}}(\calF_{N,M}))]\rightarrow 1$ as
$N$ 
goes to infinity (if such a function exists).
\end{enumerate}
Similarly, we also define $R_{\operatorname{worst}}(\calF_{N,M})$, $R_{\operatorname{best}}(\calF_{N,M})$
and $R_{\operatorname{almost}}(\calF_{N,M})$
for the randomized case.
Our results are summarized in Table~\ref{tab:ResultSummary}.
More precise description is as follows.
\begin{description}
\item[(i)] For every $1 \le M \le 2^{N/(\log{N})^{2+\epsilon}}$ with an arbitrary
small positive constant $\epsilon$,
\[
Q_{\operatorname{worst}}(\calF _{N,M}) = \Theta\left(\sqrt{N
  \frac{\log{M}}{c+\log{N}-\log\log{M}}}+\sqrt{N}\right),
\]
where $c$ is a positive constant
(Strictly speaking, the lower
bound of $Q_{\operatorname{worst}}(\calF _{N,M})$ holds for broader range $1 \le M \le 2^N/2$).
For every $1\le M \leq  2^{N-1}$, 
\[
R_{\operatorname{worst}}(\calF _{N,M}) = \Theta\left(N\right).
\]

\item[(ii)] For every $1\le M \leq  2^{N}/2$,
\[
Q_{\operatorname{best}}(\calF _{N,M}) = \Omega(\sqrt{N-\log M}),
\]
\[
R_{\operatorname{best}}(\calF _{N,M}) = \Omega(N-\log M).
\]
(For every $1\le M\le 2^{c  N}$ with an arbitrary large constant $c<1$,
the bound is optimal. In the case of  $M=2^{(1-o(1))N}$, the bound is optimal if 
$M$ is a power of 2.)
\item[(iii)] For every  $1\le M \le 2^N/2$,
\[
Q_{\operatorname{almost}}(\calF _{N,M}) =
\Theta\left(\frac{\log{M}}{c + \log{N} -  \log\log{M}} + \sqrt{N}\right),
\]
\[
R_{\operatorname{almost}}(\calF _{N,M}) =
\Theta\left(N\right),
\]
where $c$ is a positive constant.
The proof essentially implies that
$Q_{\operatorname{almost}}(\calF _{N,M})$ 
is equal to 
the average quantum query complexity $Q_{\operatorname{avg}}(\calF _{N,M})$
over uniformly distributed functions in $\calF_{N,M}$ up to a constant factor,
since the fraction of functions whose  quantum 
query complexity is not included by $Q_{\operatorname{almost}}(\calF _{N,M})$ 
is $o(1/N^k)$ for some large positive constant $k$.
Similarly, 
$R_{\operatorname{almost}}(\calF _{N,M})$ is essentially the same,
up to a constant factor, as 
the average randomized query complexity $R_{\operatorname{avg}}(\calF _{N,M})$
over uniformly distributed functions in $\calF_{N,M}$.

\end{description}
\paragraph{Implications of Our Results}
\begin{itemize}
\item Results (i) and (iii): There is a super-linear gap between the worst-case and average-case quantum query complexities if $M$ is in the range that is upper-bounded by $2^{o(N)}$ and lower-bounded by $N^{\omega (1)}$. 
The maximum gap is $\Theta(N^{3/4})$ versus $\Theta(\sqrt{N})$ at $M=2^{\sqrt{N}\log N}$.
\item Results (i) and (iii): There is no super-linear gap between 
the average-case and worst-case randomized query complexities
over $\calF _{N,M}$ for every $M$.
\item Results (i) and (ii): For every $M=O(N^{O(1)})$, any function in $\calF_{N,M}$ has quantum query complexity $\Theta (\sqrt{N})$. In other words, any function in this family
has the same quantum query complexity up to a constant factor as the OR function.
\item Results (ii): For every $M=O(2^{cN})$ 
with an arbitrary large constant $c<1$,
every function in $\calF_{N,M}$ has randomized query complexity $\Omega (N)$.
Hence, for instance, any graph property testing problem whose corresponding Boolean function has $O(2^{cN})$ 1's on its truth table 
has randomized query complexity $\Omega(n^2)$ for the number $n$ of vertices
in the bounded-error setting.
\end{itemize}

\subsection{Technical Outlines for Results (i)-(iii)} 
(i) For the quantum upper bound,
we use an algorithm~\cite{AmbIwaKawRayYam07TCS} for the Oracle Identification Problem (OIP),
which is defined as follows:
If we are given an oracle $x$ and a set $S$ of $M$ oracle candidates out of $2^{N}$ ones,
determine which oracle in $S$ is identical to $x$ with the promise
that $x$ is a member of $S$. More concretely, we set $S$ to the on-set of $f_N$,
run the algorithm, and finally verify with Grover search 
that the output of the algorithm is equal to the given $N$ bits.
To achieve the tight bound, 
we modify the algorithm so that it can work for a wider range of $M$.
For the lower bound, we give a function with on-set size $M$ for every $M$,
and prove that the lower bound of its quantum query complexity matches
the upper bound by using the quantum adversary method~\cite{Amb02JCSS}.
The lower bound of the randomized query complexity of the same function
can be proved to be $\Omega(N)$ by the classical adversary method~\cite{Aar06SICOMP}.

\noindent
(ii) The upper bound is shown by giving a function with on-set size $M$
whose quantum and randomized query complexities are $O(\sqrt{N-\log M})$
and $O(N-\log M)$, respectively.
The lower bound is proved by combining 
the edge-isoperimetric inequality on a Boolean cube and
$Q(f_N)=\Omega(\sqrt{s(f_N)})$~\cite{BeaBuhCleMosWol01JACM}
and $R(f_N)=\Omega(s(f_N))$~\cite{Nis91SICOMP}, where $s(f_N)$ is the
sensitivity of $f_N$.

\noindent
(iii) For the quantum upper bound, we encode the given $N$-bit string  $x\in \{0,1\}^N$ as
a quantum state $\ket{\psi_x}^{\otimes m}$ for some $m$ so that, for almost all Boolean
functions in $\mathcal{F}_{N,M}$,
$\ket{\psi_x}$ and $\ket{\psi_y}$ have small inner product
for every $x,y\in f^{-1}_N(1)$ with $x\neq y$.
We then perform state discrimination procedure~\cite{HarrowWinter06ARXIV} using
$\ket{\psi_x}^{\otimes m}$
to test if $x$
is
in the on-set of $f_N$, and verify the result with Grover search.
More concretely, 
let $\ket{\psi_x} = (1/\sqrt{N})\sum_{i=1}^{N} (-1)^{x_i}\ket{i}$ for $x 
 \in \{0,1\}^N$.
We prove that,
 for almost all 
 Boolean functions $f_N \in \mathcal{F}_{N,M}$,
it holds that  
 $|\bracket{\psi_x}{\psi_y}| \le 2 \sqrt{\log{M}/{N}}$
for every two different states $\ket{\psi_x}$ and 
 $\ket{\psi_y}$ where $x,y \in f^{-1}_N(1)$. 
Here, the number $m$ of the copies of $\ket{\psi_x}$ is set to
$O(\frac{\log{M}}{c+\log{N}-\log\log{M}})$~\cite{HarrowWinter06ARXIV}.
For the quantum lower bound, we use the following facts.
(1)
The number of functions in ${\mathcal{F}}_{N,M}$ is 
 $\binom{2^N}{M}$.
(2) The number of 
 Boolean functions computable with success probability more than $1/2$ 
with at most $d/2$ queries is at most $T(N,d) = 2 
 \sum_{i=0}^{D-1}\binom{2^N-1}{i}$ for $D =
 \sum_{i=0}^{d}\binom{N}{i}$ ~\cite{MonNisRay08ISAAC,BuhVerWol07CCC}.
We then calculate 
the largest $d$
such that
$T(N,d)/\binom{2^N}{M}\rightarrow 0$ for $N\rightarrow \infty$.
The randomized lower bound is lower-bounded by
the above quantum lower bound and the randomized lower bound in (ii),
from which the bound follows.

\subsection{Organization}
Section~2 defines the oracle (or black-box) model, and then gives a technical lemma and known lower bound theorems
that are used in the proofs in the following sections.  Sections~3, 4
and 5 prove the best-case, worst-case, and average-case
complexities, respectively, over family $\calF_{N,M}$. Some applications to graph property testing
are described at the end of Section~4.
Section 6 concludes the paper.


\section{Preliminaries}
We assume
the oracle (or black-box) model.
In this model,
an input (i.e., a problem instance) is given as an oracle.
For any input $x=(x_1,\ldots,x_{N})\in \{0,1\}^N$, 
we can get $x_i$ by making a query with index $i$ to the oracle.
The {\it randomized query complexity} of a
problem $P$ whose input is given as an $N$-bit string is defined as
the number of queries needed to solve $P$ with bounded-error,
i.e., with success probability at least $1/2+c$ for a constant $c>0$.
In the quantum setting, we can get a superposition of answers
by making a query with the same superposition of indices.
More formally, a unitary operator $O$,
corresponding to a single query to an oracle, 
maps $|i\rangle|b\rangle|w\rangle$ to 
$|i\rangle|b\oplus x_i\rangle|w\rangle$ 
for each $i\in [N]=\{1,2,\ldots,N\}$ and $b\in\{0,1\}$, where $w$ denotes workspace. 
A {\it quantum computation} of the oracle model (first formulated in~\cite{BeaBuhCleMosWol01JACM})
is a sequence of unitary transformations $U_{0} \to O \to U_{1} \to O \to \cdots \to O \to U_{t}$,
where $U_{j}$ is a  unitary transformation that does not depend on the input. 
The above computation sequence involves $t$ oracle calls, which is our
measure of the complexity: The {\it quantum query complexity} $Q(P)$ of a
problem $P$ whose input is given as an $N$-bit string is defined as
the number of quantum queries needed to solve $P$ with bounded-error.

This paper considers the problem of evaluating the value ($0$ or $1$)
of a Boolean function  $f(x_1,\ldots,x_N)$ over $N$ variables,
assuming that the truth table of $f$ is known.
The {\em on-set} of $f$ is the set of assignments $(x_1,\ldots,x_N)$
with $f(x_1,\ldots,x_N)=1$. We denote  by $\calF _{N,M}$ the family of all $N$-variable Boolean
functions whose on-set sizes are $M$.

In the following, we present a technical lemma
for precise analysis, and a standard lower bound theorem,
the \emph{adversary method}. The technical lemma,
together with well-known inequality $\binom{N}{k}\le \left( \frac{eN}{k} \right)^k$,
essentially gives a precise upper bound of $k$
that satisfies $\binom{N}{k}\le M$. The lemma will be used 
in the worst- and average-case analyses (i.e., Sections 4 and 5).
We assume hereafter that the base of the logarithm is $2$ when we do not explicitly write the base. 
\begin{lemma}\label{defd}
For $ 1 < z \le 2^N$, let $d(z) = \frac{\log{z}}{4\left(\log{eN} - 
 \log\log{z}\right)}$, where $e$ is the base of the natural logarithm. 
 Then, it holds that $d(z)$ is monotone non-decreasing, and 
\begin{equation}\label{propd}
\left(\frac{eN}{d(z)}\right)^{d(z)} \le z. 
\end{equation} 
\end{lemma}
\begin{proof}
The monotone non-decreasing property can be easily checked since for 
 any $1 < z \le z' \le 2^N$, $d(z) \le d(z')$. The rest of the proof 
 follows from the formula below:
by taking the log of both sides
of Eq.~\ref{propd},
\begin{eqnarray*}
d(z) \log{\frac{eN}{d(z)}} &=& 
 \frac{1}{4}\frac{\log{z}}{\log{(eN)}-\log\log{z}}\log\left(\frac{eN}{\frac{1}{4}\frac{\log{z}}{\log{(eN)}-\log\log{z}}}\right)\\
&=& \frac{1}{4}\frac{\log{z}}{\log{(eN)}-\log\log{z}}
\left(\log{(eN)}-\log\log{z}+\log{4}+\log\left({\log{eN}-\log\log{z}}\right) \right)\\
&=& \frac{1}{4}\log{z}\left(1 + \frac{2 + \log{y}}{y} \right)\\
&\le& \log{z},
\end{eqnarray*}
for $y = \log{(eN)}-\log\log{z}$, where the last inequality is due to $\log{y}/y \le 1$ for $y \ge 1$.
\end{proof}

The adversary method is originally given in ~\cite{Amb02JCSS}
(the next statement is a reformulation  due to ~\cite{Aar06SICOMP}).

\begin{theorem}[Quantum adversary method~\cite{Amb02JCSS}]
\label{quantumadv} 
Let $\mathcal{A}$ $\subseteq f^{-1}\left(  0\right)$
and $\mathcal{B}\subseteq f^{-1}\left(  1\right)  $ be sets of inputs to
a Boolean function $f$. Let $R\left(  A,B\right)  \geq0$ be a real-valued function,
and for $A\in\mathcal{A}$, $B\in\mathcal{B}$, and index $i$, let
\begin{align*}
\theta\left(  A,i\right)   &  =\frac{\sum_{B^{\ast}\in\mathcal{B}~:~A\left(
i\right)  \neq B^{\ast}\left(  i\right)  }R\left(  A,B^{\ast}\right)  }%
{\sum_{B^{\ast}\in\mathcal{B}}R\left(  A,B^{\ast}\right)  },\\
\theta\left(  B,i\right)   &  =\frac{\sum_{A^{\ast}\in\mathcal{A}~:~A^{\ast
}\left(  i\right)  \neq B\left(  i\right)  }R\left(  A^{\ast},B\right)  }%
{\sum_{A^{\ast}\in\mathcal{A}}R\left(  A^{\ast},B\right)  },
\end{align*}
where $A(i)$ and $B(i)$ denote the value of the $i$th variable for
$A$ and $B$, respectively,
the denominators are all nonzero. Then the number of quantum queries
needed to evaluate $f$ 
with probability at least $9/10$
is $\Omega\left(
1/\upsilon_{\operatorname*{geom}}\right)  $, where%
\[
\upsilon_{\operatorname*{geom}}=\max_{\substack{A\in\mathcal{A},~B\in
\mathcal{B},~i~:\\R\left(  A,B\right)  >0,~A\left(  i\right)  \neq B\left(
i\right)  }}\sqrt{\theta\left(  A,i\right)  \theta\left(  B,i\right)  }.
\]
\end{theorem}
A different function of 
$\theta(  A,i)$ and $\theta (  B,i) $
 gives a randomized lower bound.
\begin{theorem}[Classical adversary method~\cite{Aar06SICOMP}]
\label{classadv}Let $\mathcal{A},\mathcal{B},R,\theta$ be the same as
in Theorem~\ref{quantumadv}. Then the number of randomized queries needed to evaluate
$f$
with probability at least $9/10$
is $\Omega\left(  1/\upsilon_{\min
}\right)  $, where%
\[
\upsilon_{\min}=\max_{\substack{A\in\mathcal{A},~B\in\mathcal{B}%
,~i~:\smallskip\,\\R\left(  A,B\right)  >0,~A\left(  i\right)  \neq B\left(
i\right)  }}\min\left\{  \theta\left(  A,i\right)  ,\theta\left(  B,i\right)
\right\}  .
\]
\end{theorem}


\section{Best-Case Analysis}
This section gives the lowest query complexity of those of 
all Boolean functions in $\calF_{N,M}$.

\begin{theorem}[Quantum Lower Bound of Any $f$ in $\calF_{N,M}$]
\label{general weaker lower}
For every $1\le M \le 2^{N-1},$ 
any $f\in \calF _{N,M}$ has quantum query complexity $\Omega(\sqrt{N-\log M})$.
\end{theorem}
\begin{proof}
We use the sensitivity argument. 
Recall that the 
sensitivity $s_x(f)$ of a Boolean function $f$ on $x\in \{ 0,1\}^N$ is
the number of variables $x_i$ such that
$f(x)\neq f(x^i)$, where $x^i$ is the string
obtained from $x$ by flipping the value of $x_i$. 
The sensitivity $s(f)$ of $f$ is the maximum of $s_x(f)$ over all $x$.
The results of Beals et al.~\cite{BeaBuhCleMosWol01JACM} implies $Q(f)=\Omega(\sqrt{s(f)})$.
We shall prove $s(f)\ge N-\log M$ for any $f$ in $\calF _{N,M}$,
from which the theorem follows.

Let $A$ be the on-set of $f$ (note that $|A|=M$).
Let $\Gamma(A)$ be the set of 
edges 
between $A$ and $\set{0,1}^N \setminus A$
of the Boolean cube $\set{0,1}^N$.
The results in~\cite{Ber67SIAP,Har64SIAP}
on the edge-isoperimetric problem on a Boolean cube states
$|\Gamma (A)|$ is minimized when $A$ is as close to a subcube as possible;
each element of $A$ 
that minimizes $|\Gamma (A)|$
has about $\log \frac{2^N}{M}$ neighbors in $\set{0,1}^N \setminus A$. More formally, it is known that:
\begin{equation}
  \label{eq:1}
  |\Gamma (A)|\ge M\log \frac{2^N}{M}.
\end{equation}

Then,
$$s(f)=\max_x s_x(f)
\ge \frac{1}{M}\sum _{x\in A} s_x(f) 
=\frac{1}{M}|\Gamma(A)|
\ge\log \frac{2^N}{M}, \text{\hspace{5mm}($\because$  Eq.\ref{eq:1})}
$$
where we use $\sum _{x\in A} s_x(f) =|\Gamma(A)|$.
Therefore,
\begin{equation*}
  Q_{\operatorname{best}}(\calF_{N,M})=\Omega (\sqrt{s(f)})=\Omega (\sqrt{N-\log M}).
\end{equation*}
This completes the proof.
\end{proof}

Since $R(f)=\Omega (s(f))$~\cite{Nis91SICOMP}, we obtain a randomized lower bound 
with a similar argument.

\begin{theorem}[Randomized Lower Bound of Any $f$ in $\calF_{N,M}$]
\label{RandomizedLowerBound}
For every $1\le M \le 2^{N-1},$ 
any $f\in \calF _{N,M}$ has randomized query complexity $\Omega(N-\log M)$.
\end{theorem}

The next theorem shows the tightness of the above lower bounds
(note that, for $M=2^{c N}$ with any constant $c <1$, the randomized lower bound  in Theorem~\ref{RandomizedLowerBound}
is obviously tight).
\begin{theorem}[Tightness of Lower Bounds]
  For every $1\le M \le 2^{c N}$ with an arbitrary large constant $c <1$, there is a function whose quantum query complexity is $O(\sqrt{N})$.
For $M = 2^{(1-o(1))N}$, there is a function whose quantum and randomized query complexities are $O(\sqrt{N-\log M})$ and $O(N-\log M)$, respectively, if $M$ is a power of 2.
\end{theorem}
\begin{proof}
Let $C_M$ be the set of $N$-bit strings
\[
\set{0,1}^{\floor{\log M}}0^{N-\floor{\log M}},
\]
a maximal Boolean cube of size at most $M$.
Consider the function whose onset is 
\begin{equation*}
  \begin{cases}
  C_{M}& \text{if $M$ is a power of 2,}\\
  C_{M} \cup 
\left\{
y\colon y\in [0,\Delta M-1]
\right\} 
\,
10^{N-\floor{\log M}-1}
& \text{otherwise,}
  \end{cases}
\end{equation*}
where 
$\Delta M \equiv M-2^{\floor{\log M}}$ and
$y$ is a ${\floor{\log M}}$-bit string.

Suppose $M$ is a power of 2.
To evaluate this function, we first test if string ``$x_{\floor{\log M}+1}\dots x_N$'' is $0^{N-\floor{\log M}}$ with Grover's search algorithm.
If the test is passed, output $f=1$; otherwise output $f=0$.
Clearly, the quantum query complexity of this test is $O(\sqrt{N-\log M})$.

Suppose $M$ is not a power of 2.  We perform another test if the above
test is not passed.  The additional test is to check if string
``$x_{\floor{\log M}+1}\dots x_N$'' is ``$10^{N-\floor {\log M}-1}$''
and if the integer represented by $x_1\dots x_{\floor{\log M}}$ is at
most $M-2^{\floor{\log M}}-1$.  We claim that this test can be done
with $O(\sqrt{N-\log M}+\sqrt{\log M})$ quantum query complexity.  Therefore,
the overall quantum query complexity is $O(\sqrt{N-\log M}+\sqrt{\log
  M})=O(\sqrt{N})$.  

We now prove the claim.  The checking if
$w:=x_{\floor{\log M}+1}\dots x_N$ 
is ``$10^{N-\floor {\log M}-1}$''
can be done with Grover search
over 
$w\oplus 10^{N-\floor{\log M}-1}$, where $\oplus$ is bit-wise XOR,
which needs $O(\sqrt{N-\log M})$ quantum queries.
For checking if the integer represented by $z:=x_1\dots x_{\floor{\log M}}$
is at
most $M-2^{\floor{\log M}}-1$,
we just need to search the
bit $x_i$ with the largest index $i$ such that $x_i$ does not agree to
the $i$th bit of $(M-2^{\floor{\log M}}-1)_2$, where $(k)_2$ is the 
$(\floor{\log M})$-bit binary expression of integer $k$. To do this, 
we perform binary search over $z$
with Grover search. Namely,
let $\tilde{z}:=z\oplus (M-2^{\floor{\log M}}-1)_2$,
and run Grover search over the first half of $\tilde{z}$.
If no ``1''  is found, then run Grover search over the first half of the rest;
otherwise the first quarter of $\tilde{z}$. 
This procedure is recursively performed until the size of search space is at most some constant.
To bound the total error probability by some constant, 
we repeat the $k$th search $O(k)$ times.
Then the sum of error probability of each recursion is a geometric series; it is bounded by some constant.
Since the $k$th search space is of size $|\tilde{z}|/2^{k}$,
the query complexity of the $k$th search is bounded by $O(k\sqrt{|\tilde{z}|/2^{k}})$.
Therefore the quantum query complexity of the search over $\tilde{z}$ 
is the sum of $O(k\sqrt{|\tilde{z}|/2^{k}})$ over all $k$, i.e.,
$O(\sqrt{|\tilde{z}|})=O(\sqrt{\log M})$.

The randomized upper bound is obtained by a similar argument except that
sequential classical queries are used instead of Grover search.
\end{proof}


\section{Worst-Case Analysis}
In this section, we consider
the highest quantum query complexities over all Boolean functions in $\calF _{N,M}$.

To prove the upper bound, we reduce the problem to 
Oracle Identification Problem (OIP) ~\cite{AmbIwaKawMasPutYam04STACS,AmbIwaKawRayYam07TCS}
defined as follows:
Given an oracle $x$ and a set $S$ of $M$ oracle candidates out of $2^{N}$ ones,
determine which oracle in $S$ is identical to $x$ with the promise that $x$ is a member of $S$.
OIP can be solved with a constant success probability
by making $O(\sqrt{N (1+\frac{\log{M}}{\log{N}})})$ quantum queries to the given oracle
if $1 \le M \le 2^{N^d}$ for some constant $d$ $(0 < d < 1)$ ~\cite{AmbIwaKawRayYam07TCS}.
In the proof below, we 
improve the previous algorithm~\cite{AmbIwaKawRayYam07TCS} 
so that it can optimally work for a wider range of $M$,
and apply it.

Now, we give an upper bound for the query complexities 
of all Boolean functions in $\calF_{N,M}$. 
\begin{theorem}[Quantum Upper Bound of Any $f$ in $\calF_{N,M}$]\label{general upper}
\sloppy
For every $1 \le  M \le 2^{N/(\log{N})^{2+\epsilon}}$ for an arbitrary small
positive  constant $\epsilon$, any Boolean function $f \in {\cal 
 F}_{N,M}$ has quantum query complexity $O\left(\sqrt{N 
 \frac{\log{M}}{\log{N}-\log\log{M}}}+\sqrt{N}\right)$. 
\end{theorem}
\begin{proof}
We set candidate set $S$ of OIP to the on-set of $f$,
which can be constructed from the known truth table of $f$.
Note that $|S|=M$ since $f\in \calF _{N,M}$.
We then invoke the OIP algorithm~\cite{AmbIwaKawRayYam07TCS} with $S$ to find the hidden oracle
with $O (\sqrt{N (1+\frac{\log{M}}{\log{N}})})$ queries,
assuming the promise that the current oracle $x$ is in $S$ 
(actually, the promise does not hold if $f(x)=0$). 
Let $z\in\{0,1\}^N$ be the string that the OIP algorithm outputs.

If $f(x) =1$, the promise of the above OIP is indeed satisfied;
$z$ is equal to $x$ with high probability.

If $f(x) = 0$, the promise 
does not hold; the OIP algorithm outputs some answer $z\in S$ (note that $z\neq x$). 
To recognize this case, it suffices to check whether $z$ is equal to $x$ by using
Grover search with $O (\sqrt{N})(\in O (\sqrt{N (1+\frac{\log{M}}{\log{N}})})) $ queries.
This completes the proof for $1\le M \le 2^{N^d}$ 
for some constant $k$ and any constant $0<d<1$.

For bigger $M$, we cannot use the original OIP
algorithm~\cite{AmbIwaKawRayYam07TCS}.  Very roughly speaking, the OIP
algorithm recursively repeats the following procedure.  
Suppose that the given candidate set is represented by an $M$-by-$N$ matrix,
in which each row corresponds to a candidate.
First collect
the set $T$ of columns each of which covers
(i.e., has 1 at the positions of)
a disjoint fraction that is at least  $\beta $ and  at most some constant, of
the current rows (candidate) set $S$, and then apply Grover search to the oracle
restricted to set $T$  to find 1; if 1 is found, we can reduce the row set into the 
fraction.
For small $\beta$, we may reduce the candidates into a small set of
rows, but the Grover search may cost too much since the cardinality of $T$
can be  roughly $1/\beta$; $\beta$ must be set to an appropriate value~\cite{AmbIwaKawRayYam07TCS}:
\[
 \beta = (\log{M} (\log\log{M})^2 \log{N})/(2N).
\]
If the Grover search fails, 
the rows covered by $T$ are excluded from the matrix, and the remaining matrix is sparse. To further reduce the set rows of the sparse matrix, multi-target Grover search~\cite{BoyBraHoyTap98FP} is used with promise that the fraction of 1 over $N$ bits in the oracle is $\gamma<1/2$. 
The proof in \cite{AmbIwaKawRayYam07TCS} shows that
if $\gamma$ is adjusted to 
$
\Theta({\log{|S|}}/{(N\log{N})})
$
so that
the number of $N$ bit strings with Hamming weight at most $\gamma N$ is 
about the square root of $|S|$, the total query complexity is
$O (\sqrt{N \log{M} \log{N}}/\log{(1/\beta)})$, which gives
$O (\sqrt{N \log{M}/ \log{N}})$ for 
$M\le 2^{N^d}$.

To expand the range of $M$ for which the algorithm can work,
we slightly decrease the value of $\beta$ to handle large $M$:
\[
 \beta^{\prime} = (\log{M} (\log\log{M})^2 \log{(eN /\log{M})})/(2N).
\]
To meet $\beta^{\prime} < 1$,  it is required that  $M \le 2^{N/(\log{N})^{2+\epsilon}}$.
Note that $\beta^{\prime} =\Theta (\beta)$ for the original range of $M$, $M\le 2^{N^d}$.
We can also set $\gamma$ to a more precise value 
satisfying $\sum_{k=0}^{\gamma N}\binom{N}{k} \le |S|^{1/2}$
by virtue of Lemma~\ref{defd}, namely, 
$$
\gamma^{\prime} = \Theta\left(\frac{\log{|S|}}{N(\log{eN}-\log\log{|S|})}\right).
$$
These changes of parameter values yield
the total query complexity of 
\[
O(\sqrt{N \log{M} \log{(eN /\log{M})}}/\log{(1/\beta^{\prime})}),
\]
which gives the complexity in the statement.
The details of the proof are the same with those in the original algorithm~\cite{AmbIwaKawRayYam07TCS}.
\end{proof}

The following corollary is immediate.

\begin{corollary}\label{cor0621}
For every
$M= O(N^{O(1)})$,
any function $f\in \calF_{N,M}$ has quantum query complexity $O (\sqrt{N})$.
\end{corollary}

The following theorem shows 
the bound in Theorem~\ref{general upper} is tight.

\begin{theorem}[Tightness of the Upper Bound]\label{general lower}
For every  $1 \le M \le 2^{N-1}$,  there is a function $f\in {\cal{F}}_M$ 
whose quantum and randomized query complexities are $\Omega\left(\sqrt{N \frac{\log{M}}{c + \log{N}-\log\log{M}}}+\sqrt{N}\right)$ 
for a positive constant $c$ and $\Omega(N)$, respectively.
\end{theorem}
\begin{proof}\sloppy
If $1\le M \le N^2$, the upper bound $O(\sqrt{N})$ given in Theorem~\ref{general upper} matches the lower bound given in Theorem~\ref{general weaker lower}.
This implies that there exists a function with query complexity
 $\Omega\left(\sqrt{N \frac{\log{M}}{c + \log{N}-\log\log{M}}}+\sqrt{N}\right)=\Omega (\sqrt{N})$.

Suppose that $N^2\le M \le 2^{N-1}$.
Let $k$ be the integer that satisfies
$$
D = \sum_{i=0}^{k}\binom{N}{i} \le M \ \ \mbox{and}\ \  \sum_{i=0}^{k+1}\binom{N}{i} > M.
$$
Consider a Boolean function $f$ such that 
$f(x)=1$ for all $x$ with $\mbox{Ham}(x)\leq k$ and for $M-D$ assignments $x$ with $\mbox{Ham}(x)\ge k+2$, 
and $f(x) = 0$ for all the remaining assignments. Here, $\mbox{Ham}(x)$ denotes the Hamming weight of $x$. 
We claim that the quantum query complexity of $f$ is $\Omega(\sqrt{Nk})$, which is proved later. 
To complete the proof, it suffices to show that 
$k = \Omega(d(M/N))$ for the function $d(\cdot)$ defined in Lemma~\ref{defd},
since
 $\Omega(d(M/N)) = \Omega(\log{M}/(\log{eN}-\log\log{N}))$:
by simple algebra, it holds that (by assuming $d(M/N)$ is an integer for simplicity):
$$
\sum_{i=0}^{d(M/N)}\binom{N}{i} \le N\binom{N}{d(M/N)} \leq N\left(\frac{eN}{d(M/N)}\right)^{d(M/N)} \le M,
$$
where the last inequality is due to Lemma~\ref{defd}.

Now we prove the claim.
Let ${\cal A} \subseteq f^{-1}(0)$ and ${\cal B} \subseteq f^{-1}(1)$ 
be defined as the sets of $x$'s with $\mbox{Ham}(x)=k+1$ and $\mbox{Ham}(x) =k$, respectively.
For any $A \in {\cal A}$ and $B \in {\cal B}$, let us define the relation $R$ in Theorem~\ref{quantumadv} 
as $R(A,B) = 1$ if $A$ and $B$ differ in exactly one position
and $R(A,B) = 0$ otherwise.
Then, it can be shown that $\theta(A,i)= 1/(k+1)$, and $\theta(B,i)= 1/(N-k)$
by the definition of $A$ and $B$.
Hence we obtain the lower bound $\Omega (\sqrt{Nk})$ by Theorem~\ref{quantumadv}.  

The randomized lower bound is obtained by applying Theorem~\ref{classadv} with a similar argument.
\end{proof}
\begin{remark}
The proof of Theorem~\ref{general upper} shows that
computing $f\in \calF_{N,M}$ is  reducible to OIP with $M$ candidates and Grover search.
Since Grover search has query complexity  $O(\sqrt{N})$,
Theorem~\ref{general lower} implies 
 $\Omega\left(\sqrt{N \frac{\log{M}}{c + \log{N}-\log\log{M}}}\right)$
is also a lower bound of OIP
for every $N^{\omega(1)}<M\le 2^N/2$.
(The query complexity of OIP is $\Omega(N)$ 
for every $2^N/2<M\le 2^N$, since
OIP with $M$ candidates is reducible to OIP with $M^{\prime}\ (>M)$ candidates.)
This is an improvement over the lower bound
of OIP in~\cite{AmbIwaKawRayYam07TCS} for large $M$.
\end{remark}

\subsection*{Applications}
As an application of Theorem~\ref{general upper},
we consider the problem of graph property testing, i.e., the
problem of testing  if $G$ 
 has a certain property for a given graph $G$.
More precisely, an $n$-vertex graph is given as $n(n-1)/2$ Boolean
variables, $x_{i}$ for $i\in \{ 1,\dots , n(n-1)/2\}$, representing
the existence of the $i$th possible edge $e_i$, i.e., 
$x_{i} =1$ if and only
if  $e_i$ exists.  In this setting,
graph property testing is just the problem of evaluating a Boolean function $f$
depending on the $n(n-1)/2$ variables such that 
$f(x_1,\dots , x_{n(n-1)/2})=1$
if and only if the
graph has a certain property.
An interpretation of graph property testing according to Theorem~\ref{general upper} is to decide if $G$ is a member of $\mathcal{F}$ for the family $\mathcal{F}$ of all graphs with certain properties.  
Thus, Theorem~\ref{general upper} directly gives the next lemma
with $M=|\mathcal{F}|$ and $N=n(n-1)/2$.
  \begin{lemma}
\label{lm:quantum upper bound of graph property}
Any graph property $P$
can be tested  with 
$O\left(\sqrt{n^2\frac{\log|\mathcal{F}|}{c+\log n- \log\log |\mathcal{F}|}}+n\right)$ 
quantum queries for a positive constant $c$, 
where $\mathcal{F}$ is the family of all graphs having property $P$,
if $1\le |\mathcal{F}|\le
    2^{\binom{n}{2}/{(\log \binom{n}{2})^{2+\epsilon}}}$ for an arbitrary small positive constant $\epsilon$.
  \end{lemma}
An interesting special case is graph isomorphism testing against a fixed graph,
the problem of deciding if a given graph $G$
is isomorphic to 
an arbitrary fixed graph $G'$.
\begin{theorem}[Graph Isomorphism Testing against a Fixed Graph]
  Graph isomorphism testing against a fixed graph has $O(n^{1.5})$ quantum query complexity
and $\Omega(n^2)$ randomized query complexity.
\end{theorem}
\begin{proof}
  The number of graphs isomorphic to $G'$ is at most the number of
  permutations over the vertex set, i.e., $n!=2^{O(n\log n)}$,
from which together with Lemma~\ref{lm:quantum upper bound of graph property}
the quantum upper bound follows.
The randomized lower bound follows from Theorem~\ref{RandomizedLowerBound}.
\end{proof}
This upper bound is optimal in the worst case over all possible $G'$, since
the lower bound $\Omega (n^{1.5})$ 
of connectivity testing problem in Ref.~\cite{DurHeiHoyMha06SICOMP}
is essentially the lower bound of deciding whether
a given graph is isomorphic to one cycle or two cycles.

Another interesting special case is \emph{graph genus testing}, the problem
of testing if a given graph is a connected graph with genus $g$.  Informally, the
\emph{genus} of a connected graph $G$ is the minimum number of handles that need to be
added to the plane so that the graph can be drawn without edge crossing (see, e.g., \cite{Diestel00GraphBook}).
Note that for $g=0$, graph genus testing is planarity testing, i.e.,
determining if a given graph is planar.

\begin{theorem}[Graph Genus Testing]
For $g=\left\{\binom{n}{2}\right\}^c=O(n^{2c})$ for an arbitrary large constant $0\le c<1$,  graph genus testing has $O(n\sqrt{n+g})$ quantum query complexity and $\Omega(n^2)$ randomized query complexity.
\end{theorem}
\begin{proof}
For any connected graph embedded on a  surface of genus $g$,
Euler's equation (see, e.g., ~\cite{Diestel00GraphBook}) says $n - m + f = 2 - 2g$,
where   $n\ge 3$, $m$ and $f$ are the numbers of vertices, edges and faces.
Every face is adjacent to at least three edges
and every edge is adjacent to at most two faces, from which we have $f\le 2m/3$.
Hence, $m\le 3({n-2+2g})$,
and then
$|\calF|
\le 
\sum _{i=0}^{m}\binom{n^2}{i}
\le (m+1)n^{2m}=2^{O(m\log n)}
=2^{O((n+g)\log n)}.
$
Since $g=\left\{\binom{n}{2}\right\}^c$ for constant $0\le c<1$,
$|\calF|\le 2^{O((n+g)\log n)}< 2^{\binom{n}{2}/{(\log \binom{n}{2})^{2+\epsilon}}}$
for sufficiently large $n$. Therefore, we can apply Lemma~\ref{lm:quantum upper bound of graph property} to obtain the quantum upper bound.
The randomized lower bound is due to 
Theorem~\ref{RandomizedLowerBound}.
\end{proof}


\section{Average-Case Analysis}
This section considers 
the upper and lower bounds for the quantum query complexities 
of almost all functions in $\calF _{N,M}$. To prove the upper bound, 
we need the following lemmas.  
The first one bounds the inner product of two quantum states associated with two different oracles.
The second one is a result of quantum state discrimination.

\begin{lemma}\label{randomInstance}
Let $\ket{\psi_x} = \frac{1}{\sqrt{N}}\sum_{i=1}^{N} (-1)^{x_i}\ket{i}$ for $x=(x_1,\dots, x_N) \in \{0,1\}^N$. 
For any $f$ in at least $(1-2/M^{0.88})$ fraction of 
${\cal F}_{N,M}$ 
with $N\le M \le 2^{N-1}$,
it holds that $|\bracket{\psi_x}{\psi_y}| \le 2 \sqrt{\frac{\log{M}}{{N}}}$ 
for every two different states $\ket{\psi_x}$ and $\ket{\psi_y}$ where $x,y \in f^{-1}(1)$.
\end{lemma}
\begin{proof}
Since $|\bracket{\psi_x}{\psi_y}| \le 1$ obviously holds for every two quantum states, 
we will only show the lemma when $N\le M \le 2^{N/4}$. Notice that by the definition, 
    \begin{eqnarray*}
\langle \psi _x|\psi _y\rangle
&=& \frac{1}{N} \sum _{i=1}^N (-1)^{x_i\oplus y_i}\\
&=& \frac{1}{N} \sum _{i=1}^N (1 - 2(x_i\oplus y_i))\\
&=& \frac{1}{N} (N - 2\Ham(x,y)),
    \end{eqnarray*}
where $\Ham(x,y)$ is the Hamming distance of $x$ and $y$.

We can prove the following claim (The proof can be found in Appendix).
\begin{claim}
\label{cl:AverageHammingBound}
If $f$ is uniformly distributed over ${\cal F}_{N,M}$ with $M \le 2^{N/4}$, 
then 
$\Ham (x,y) \ge N\left(\frac{1}{2} - \sqrt{\frac{(2+\epsilon)}{\log e}\frac{\log M}{2N}}\right)$ 
holds 
for every pair of different $x,y\in f^{-1}(1)$
with probability $1-2/M^\epsilon$, where $\epsilon$ is an any positive constant. 
\end{claim}
The lemma then follows from the claim 
by setting $\epsilon=2\log e-2>0.88$.
\end{proof}
  \begin{lemma}[\cite{HarrowWinter06ARXIV}]
\label{lm:QuantumStateDiscrimination}
Suppose that a set of $M$ quantum states, $\{\ket{\phi_x}\}_{x\in S}$,
is known, where
$S$ is an index set of cardinality $M$, and that
  $|\bracket{\phi_x}{\phi_y}|^2 \le F<1$ for any pair of different $x, y\in S$.  
If $m =
  O(\log{M}/\log{(1/F)})$ copies of unknown $\ket{\phi_x}$, i.e., $\ket{\phi_x}^{\otimes m}$, are
  given, it is possible to identify index $x$ with probability at least
  $2/3$.
  \end{lemma}

Now, we are ready to show an upper bound for the quantum query complexities 
of almost all functions in $\calF_{N,M}$. 

\begin{theorem}[Quantum Upper Bound for Almost All $f$ in $\calF_{N,M}$]\label{quantum almost-all upper}
\sloppy
For every $1 \le M \le 2^{N-1}$, 
any Boolean function in
at least $(1-1/N^k)$ fraction of
${\cal F}_{N,M}$ has quantum query complexity $O(\frac{\log{M}}{c + \log{N} - \log\log{M}} + \sqrt{N})$,
where $k\ge 1$ is an arbitrary constant and $c$ is a certain positive constant.
\end{theorem}
\begin{proof}
If $1\le M \le N^d$ for an arbitrary constant $d\ge 2$, all functions in $\calF_{N,M}$ has query complexity
$\Theta (\sqrt{N})$ by Corollary~\ref{cor0621}; the theorem holds.
If $M>2^{N/5}$, we will prove the lower bound is $\Omega (N)$ in Theorem~\ref{quantum almost all lower}.

Suppose $N^d<M\le2^{N/5}$.
We give an algorithm for computing $f$ based on 
Lemma~\ref{lm:QuantumStateDiscrimination}.
Set $S:=f^{-1}(1)$.
We then  create
  $m$ copies of quantum state $\ket{\psi_x}$, each of which requires only one query,
where $x$ is the $N$-bit string in the given oracle.
Lemma~\ref{randomInstance} says that
$|\bracket{\psi_x}{\psi_y}| \le 2 \sqrt{\frac{\log{M}}{{N}}}$ 
for any $f$ in at least $(1-2/N^{0.88d})$ fraction of 
${\cal F}_{N,M}$.
Suppose $f$ is in the fraction.
By setting $F=4\frac{\log{M}}{{N}}$, Lemma~\ref{lm:QuantumStateDiscrimination}
says that we can identify $x$ 
with only $m =
  O(\log{M}/(c + \log{N} - \log\log{M}))$ copies, i.e., with only $m$ queries
with probability at least 2/3, if $x\in S$.

If $x\not\in S$, the output may be some $y\in S$ (obviously, $x\neq y$).
This case can be detected by running Grover search over $x\oplus y$, where $\oplus$ is
bit-wise XOR.

In summary, our algorithm first performs the quantum state discrimination procedure to identify $x$,
and then runs Grover search to test if the output of the above procedure is equal to $x$.
The total quantum query complexity is 
$O(\frac{\log{M}}{c + \log{N} - \log\log{M}} + \sqrt{N})$.
We can set $d\ (\ge 2)$ such that $2/N^{0.88d} \le 1/N^k$ for every $k\ge 1$. Therefore, the theorem follows.
\end{proof}

We can show the optimality of Theorem~\ref{quantum almost-all upper} as follows.

\begin{theorem}[Quantum Lower Bound for Almost All $f$ in $\calF_{N,M}$]\label{quantum almost all lower}
For every $1 \le  M \le 2^{N-1}$, at least $1-1/2^N$ fraction of ${\cal{F}}_{N,M}$ have quantum query 
 complexity
$\Omega(\frac{\log{M}}{c + \log{N} - \log\log{M}} + \sqrt{N})$,
where $c>0$ is a certain constant.
\end{theorem}
\begin{proof}
If $1\le M \le 2^{\sqrt{N}}$,
the query complexity of all Boolean functions in $\calF_{N,M}$ 
 is $\Omega(\sqrt{N})$
by Theorem~\ref{general weaker lower}; the theorem holds.
Thus, we shall prove the theorem for $2^{\sqrt{N}} < M \le 2^{N-1}$ 

We shall bound the number of quantum queries  by the 
 monotone non-decreasing function $d(z)$ in Lemma~\ref{defd}.
First, notice that the number of functions in ${\cal{F}}_{N,M}$ is
 $\binom{2^N}{M}$, which is at least $\left(\frac{2^N}{M}\right)^M =
 2^{M'}$ for $M' = M(N-\log{M})$.  Secondly, notice that the number
 of Boolean functions computable with success probability more than
 $1/2$ with at most $d/2$ queries is at most $T(N,d) = 2
 \sum_{i=0}^{D-1}\binom{2^N-1}{i}$ for $D =
 \sum_{i=0}^{d}\binom{N}{i}$.  This bound is derived from the
 following two properties of a {\em sign-representing polynomial} $p$,
 a real-valued polynomial with properties that $p(x)$ is positive
 whenever $f(x) = 0$ and $p(x)$ is negative whenever $f(x) = 1$: (i)
 The {\em unbounded-error} quantum query complexity of a Boolean function $f$,
 where the success probability is only guaranteed to be more than
 $1/2$, is exactly half of the minimum degree of its sign-representing
 polynomial \cite{MonNisRay08ISAAC,BuhVerWol07CCC}.  (ii) The number of Boolean
 functions whose minimum degrees of sign-representing polynomials are
 at most $d$ is $T(N,d)$~\cite{Ant95DAM}.

We shall complete the proof of the theorem by the following three claims. 
Claims~\ref{cl1} and \ref{cl2} show that, for $z =\frac{M'}{(N+1)^2}$, 
the value of $T(N,d(z))$ (or, the number of functions computable with quantum queries at most $d( z)/2$) 
is very small compared to $2^{M'}$, i.e., $T(N,d(z))/|{\cal{F}}_{N,M}| \le 1/2^N$. 
Claim~\ref{cl3} proves the number of queries in the theorem.

\begin{claim}\label{cl1} 
For large $N$,
$T(N,d(z)) \le \frac{1}{2^N} 2^{ND}$.
\end{claim}

\begin{claim}\label{cl2} For $z = \frac{M'}{(N+1)^2}$, it 
 holds that $ND \le M'$.
\end{claim}

The theorem follows since, by Claims \ref{cl1} and \ref{cl2}, 
 $T(N,d)/|{\cal{F}}_{N,M}| \le \frac{1}{2^N} 2^{ND - M'} \le \frac{1}{2^N}$, for 
 the number of queries $d\left(\frac{M'}{(N+1)^2}\right)$ whose lower 
 bound is proved by the following claim.
\begin{claim}\label{cl3} 
$$
d\left(\frac{M'}{(N+1)^2}\right) = \Omega\left(\frac{\log{M}}{c + \log{N} - \log\log{M}}\right).
$$
\end{claim}

Below are the proofs of the claims.

\begin{proof}[Claim \ref{cl1}]
By definition of $T(N,d)$, we have
\begin{eqnarray*}
T(N,d) &=& 2 \sum_{i=0}^{D-1} \binom{2^N-1}{i}\ \leq\ 2D \binom{2^N-1}{D-1}
\ \leq\ 2D \left(\frac{e(2^N-1)}{D-1}\right)^{D}\\
&=& 2^{1 + \log{D} + D\log{e} - D\log{(D-1)} + ND}\\
&\le& \frac{1}{2^N} 2^{ND},
\end{eqnarray*}
where the last inequality is due to $2^{1 + \log{D} + D\log{e} - D \log{(D-1)}} 
 \le 2^{2D - D\log{(D-1)}}\le 1/2^D \le {1}/{2^N}$ for large $N$.
\end{proof}

\begin{proof}[Claim \ref{cl2}]
By approximating the sum of binomials, we have, for $z = \frac{M'}{(N+1)^2} \le 
 \frac{M'}{N(N+1)}$, 
\[
D = \sum_{i=0}^{d(z)}\binom{N}{i} \le 
 (d(z)+1)\left(\frac{eN}{d(z)}\right)^{d(z)}\label{eq1} \le (N+1)z \le \frac{M'}{N},
\]
where the second last inequality is due to Lemma~\ref{defd} and $d(z) 
 \le N$.
\end{proof}

\begin{proof}[Claim \ref{cl3}]
Recall that $d(z)$ is a monotone non-decreasing function, and therefore, 
 because $M' = M(N-\log{M}) \ge M$, we have 
\begin{eqnarray*}
d\left(\frac{M'}{(N+1)^2}\right) &\ge& d\left(\frac{M}{(N+1)^2}\right)
\ =\ \frac{1}{4}\frac{\log\left(\frac{M}{(N+1)^2}\right)}{\log{(eN)}-\log\log\left(\frac{M}{(N+1)^2}\right)}\\
&=& \Omega\left(\frac{\log{M}}{c + \log{N}-\log\log{M}}\right).
\end{eqnarray*}
\end{proof}
This completes the proof of Theorem~\ref{quantum almost all lower}.
\end{proof}

The above results essentially give the average quantum query complexity 
of uniformly distributed functions over $\calF_{N,M}$.
\begin{corollary}
For every $1 \le M \le 2^{N-1}$, the average quantum query complexities 
over uniformly distributed Boolean functions in
${\cal F}_{N,M}$ is $\Theta
(\frac{\log{M}}{c + \log{N} - \log\log{M}} + \sqrt{N})$,
where $c > 0$ is a certain constant.
\end{corollary}
\begin{proof}
\sloppy
By Theorems~\ref{quantum almost-all upper} and \ref{quantum almost all lower},
at least $1-(1/N^2+1/N^2)=1-2/N^2$ fraction of $\calF_{N,M}$ has query complexity
$\Theta(\frac{\log{M}}{c + \log{N} - \log\log{M}} + \sqrt{N})$.
Since the remaining fraction contributes to the average
by at most $N\cdot (2/N^2)<1$, the corollary follows.
\end{proof}

In the randomized setting, almost all functions in $\calF_{N,M}$ are hard to compute for every $M$.
\begin{theorem}[Randomized Lower Bound for Almost All $f$ in $\calF_{N,M}$]\label{RCrandomized}
For every $1 \le  M \le 2^{N-1}$, at least $1-1/2^N$ fraction of ${\cal{F}}_{N,M}$ has randomized  complexity $\Omega(N)$.
\end{theorem}
\begin{proof}
  If $M\le 2^{\epsilon N}$ for any constant $0<\epsilon<1$,
  Theorem~\ref{RandomizedLowerBound} gives $\Omega (N)$ lower bound.
  Suppose that $M=2^{(1-o(1))N}$.  Since, for every function, the
  randomized query complexity is at least the quantum query
  complexity, 
Theorem~\ref{quantum almost all lower} implies that
a lower bound of randomized query complexity is also
  $\Omega(\frac{\log{M}}{c + \log{N} - \log\log{M}})$ for at least
  $1-1/2^N$ fraction of $\calF_{N,M}$.
For $M=2^{(1-o(1))N}$, this bound is $\Omega(N)$.
This completes the proof.
\end{proof}


\section{Conclusion}
We gave the tight bounds of the worst-case, average-case, and best-case query complexities over family $\calF_{N,M}$ for every on-set size $M$ except
the upper bound of the worst-case quantum query complexity $Q_{\operatorname{worst}}(\calF_{N,M})$.
The upper bound was proved for
$1 \le M \le 2^{N/(\log{N})^{2+\epsilon}}$ with any
small positive constant $\epsilon$ and it matches the lower bound for this range of $M$.
Since we know $Q_{\operatorname{worst}}(\calF_{N,M})=\Omega (N)$ only for 
$M=\Omega (2^{cn})$ for any constant $0<c<1$,
there is still a gap between the upper and lower bounds
of $Q_{\operatorname{worst}}(\calF_{N,M})$ for 
$2^{N/(\log{N})^{2+\epsilon}}< M <2^{cn} $.
It is an open problem to close this gap.

We showed an application of the worst-case and best-case complexity bounds to some graph property testing problems. However, our bounds cannot give a good bound
for all graph property testing problems.
It would be interesting to find more problems 
to which our results can give a tight bound.


\subsection*{Acknowledgments}
The authors are grateful to Kazuyuki Amano and Mario Szegedy for valuable comments.  

\bibliographystyle{abbrv}
\bibliography{OnsetJournal}
\clearpage
\appendix
\section*{Appendix}
\begin{proof}\textit{of Claim~\ref{cl:AverageHammingBound}\\}
The probability that an element $x\in f^{-1}(1)$ has Hamming distance larger than $r$ 
from every $y$ $(\neq x)$ in $f^{-1}(1)$ is
\[
\Pr_x[\forall y \in f^{-1}(1):\ \Ham(x,y) > r]  = \frac{\binom{2^N - D}{M-1}}{\binom{2^N -1}{M-1}},
\]
where $D=\sum _{i=0}^{r} \binom{N}{i}$. Note that $D$ has the following upper bound 
as a consequence of Chernoff's inequality;
\begin{equation}\label{eq1019-1}
D < 2^N \exp\left(-2N\left(\frac{1}{2} - \frac{r}{N}\right)^2\right).
\end{equation} 

By the union bound, the probability that there is at least one $x \in f^{-1}(1)$ 
which has a neighbor $y \in f^{-1}(1)$ within the Hamming distance $r$ is 
\begin{eqnarray*}
\Pr[\exists x,y \in f^{-1}(1):\ \Ham(x,y) \le r] &\le&
\sum_{x \in f^{-1}(1)} \left(1 - \Pr_x[\forall y \in f^{-1}(1):\ \Ham(x,y) > r]\right)\\ 
&=& 
M\left(
1-\frac{\binom{2^N - D}{M-1}}{\binom{2^N -1}{M-1}}
\right).
\end{eqnarray*}
We show that this probability is $o(1)$ for some $r=r^*$, implying that
with probability $1-o(1)$, any $x,y \in f^{-1}(1)$ have Hamming distance at least $r^*$, 
as follows. Here, let $n^{\underline{m}}=\frac{m!}{(n-m)!}$.  
\begin{eqnarray*}
M\left(
1-\frac{\binom{2^N - D}{M-1}}{\binom{2^N -1}{M-1}}
\right)&=&
M\left(
1-\frac{(2^N - D)^{\underline{M-1}}}{(2^N -1)^{\underline{M-1}}}
\right)\\
&\le&
M\left(
1-\left(\frac{2^N - D-(M-1)+1}{2^N -1-(M-1)+1}\right)^{M-1}
\right)\\
&=&
M\left(
1-\left(1- \frac{D-1}{2^N -M+1}\right)^{M-1}
\right)\\
&\le &
M\left(
1-\left(1- \frac{D}{2^N/2+1}\right)^{M}
\right)\\
&\le&
M\left(
1-\left(1- \frac{MD}{2^N/2+1}\right)
\right)
\text{\hspace{1cm}(since $(1-x)^n\ge
  (1-nx)\ \mbox{for all}\ x\in (0,1)$)}\\
&=&
\frac{2M^2D}{2^N+2}\\
&\le& \frac{2M^2}{2^N+2}2^N\exp \left( 
-2N\left(
\frac{1}{2}-\frac{r}{N}
\right)^2
\right) \text{\hspace{1cm}(by Eq.(\ref{eq1019-1}))}\\
&\le& 2M^2\exp \left( 
-2N\left(
\frac{1}{2}-\frac{r}{N}
\right)^2
\right).
\end{eqnarray*}
Thus, for any positive constant $\epsilon$, if we let
\[
r=r^*
=N\left(\frac{1}{2} - \sqrt{\frac{(2+\epsilon)}{\log e}\frac{\log M}{2N}}\right),
\]
then we have 
\[
\Pr [\exists x,y \in f^{-1}(1):\ \Ham(x,y) \le r] 
 \le 2M^2  \exp \left( 
-2N\left(
\frac{1}{2}-\frac{r}{N}
\right)^2
\right)
=2/M^\epsilon.
\]

\end{proof}


\end{document}